\documentclass[preprint,3p,authoryear,12pt]{elsarticle}
\usepackage{graphicx}
\usepackage{adjustbox}
\usepackage{amssymb}
\usepackage{amsthm}
\usepackage{amsmath}
\usepackage{amsfonts}
\usepackage{tabularx}
\usepackage{array}
\usepackage{gensymb}
\usepackage{nomencl}
\usepackage{multicol}
\usepackage{lineno}
\usepackage{natbib}
\usepackage{multirow}
\usepackage{booktabs}
\usepackage{siunitx}
\usepackage{url}
\usepackage{fixltx2e}
\usepackage{pbox}
\usepackage[table,xcdraw]{xcolor}
\usepackage{array,tabularx}
\usepackage{multirow}
\usepackage{gensymb}
\biboptions{comma}
\journal{Progress in Nuclear Energy}
 
\usepackage[labelfont=bf,labelsep=period]{caption}
\usepackage{subfig}

\DeclareGraphicsExtensions{.pdf,.png}

\begin{document}

\begin{frontmatter}

\title{The challenges of gas-cooled reactor technology for space propulsion and the development of the JANUS space reactor concept}
\author[man1]{Aiden Peakman} \corref{cor1}\ead{aiden.w.peakman@uknnl.com}
\author[man1]{Robert Gregg}

\address[man1]{National Nuclear Laboratory, Chadwick House, Warrington, WA3 6AE}

\begin{abstract}

There is a strong motivation to develop high-power output nuclear fission reactors (around 1~MWe) for space applications, such as high-payload missions and long-duration missions beyond Mars, where the reduced solar flux makes using alternative energy sources challenging. Many published gas-cooled reactor designs for space applications deliver outputs in the 100~kWe regime, with typical power densities of around 4~MW/m\textsuperscript{3}. Here we present a gas-cooled reactor design -- referred to as JANUS -- employing He-Xe coolant and operating on a direct Brayton cycle, that can achieve a high power output (around 0.8~MWe) and higher power density (24~MW/m\textsuperscript{3}) than previously published designs. The core employs a coated particle fuel form with uranium nitride kernels in a graphite matrix and with a high packing fraction (45\%).
 
Starting from the CEA's published OPUS space reactor design, a variety of approaches were considered for achieving high power densities in a gas-cooled reactor, including changing the core aspect ratio. Our JANUS design uses a decreased coolant channel diameter and fuel element pitch, whilst increasing the number of fuel elements and employing radial enrichment zoning. The design achieves a significant increase in power density whilst obeying core-life limits of 2000 Effective Full Power Days and a peak fuel operating temperature of 1900~K. The core has been modelled neutronically with the ERANOS and SERPENT codes, with a simple yet robust thermal dimensioning tool developed by us to study peak fuel temperatures, to ensure the fuel operates within its design limitations and to aid optimisation.

\end{abstract}


\end{frontmatter}


\section{Introduction}
\label{sec:intro}
Most interplanetary missions rely on either solar power or decay heat from radioactive sources using radioisotope thermoelectric generators (RTGs) for electricity production \citep{LAUN-2008}. However, in order to generate sufficient power for a large mission payload, especially for long-duration missions beyond Mars, the role of a conventional nuclear reactor becomes more apparent. This is a result of reduced solar flux beyond Mars and the low power densities possible using alternative solutions. In addition, there is desire to employ electrical propulsion for long duration missions due to the improved fuel efficiency and reliability offered by ion-propulsion \citep{NOCA-2002, NASA-2004, REF02}. To deliver electricity for onboard systems including electrical propulsion, a relatively high electrical output of around 1~MWe is required for the following mission types \citep{REF02}:

\begin{itemize}
\itemsep-4pt
\item Near Earth Orbiter (NEO) deflection: a hypothetical mission to deflect large asteroids capable of threatening life on Earth by acting as a gravity tractor€™
\item Outer Solar System missions with large mission payloads between 3 to 10 tonnes
\item Lunar orbit tug: a space vehicle capable of transferring heavy mission payloads from low earth to high earth orbit
\item Cargo support for a manned Mars mission
\end{itemize}

For relatively large electrical outputs, a high core outlet temperature (around 1300~K \citep{REF12}) is also necessary to reduce launch mass and volume by enhancing thermal efficiency and ensuring the radiator size required to reject heat to space is minimised. Furthermore, to ensure the reactor design is available in the short to medium term, reliance on previous conceptual designs and tested materials is essential. For all of the above applications, there will be no human contact during operation. Therefore, in the case of a nuclear reactor being employed to power such missions, the normal criteria relating to minimising dose to humans and the environment will not be an overriding priority. 

The objective of this study was to adapt an existing reactor design to target the above applications. Therefore the reactor design must generate a sufficiently high electrical output whilst optimising both core mass and volume. Furthermore, the trade-off between increasing core power density and impacts on systems outside the core region, for example the power conversion technology are considered.

There have already been several large programmes creating conceptual closed thermal cycle nuclear energy systems with high core outlet temperatures (around 1300~K) and moderate to high electrical outputs (around 100~kWe or more). Table \ref{Tab:SpaceReactors} summarises a number of key space reactor power systems including: those operating at moderate to high electrical outputs (ERATO, FBR, the nominal OPUS concept, NPPU and SP-100); those that employ gas coolants (FBR, the nominal OPUS concept and NPPU); some of the historic designs that were launched (Topaz-1 and SNAP-10A) or extensively tested (Topaz-2); and some of the more modern designs (HOMER, KRUSTY and SAFE-400).

\begin{table}[h!]
\centering
\caption{Summary of space reactor power systems}
\begin{adjustbox}{width=1\textwidth}
\small
\begin{tabular}{|c|c|c|c|c|c|c|}
\hline
\textbf{Reactor}                &  \textbf{Country} & \textbf{Coolant}                      & \textbf{Fuel} & \textbf{Spectrum} & \textbf{Power}  &  \textbf{Reference(s)} \\ \hline
ERATO                             &  France              &  Li                                           & UN               &  Fast                     &0.02 - 0.2 MWe & \cite{REF09, REF10} \\ \hline
FBR                                  &  USA                  &  He                                         & UO$_2$       & Thermal                &$>$ 10 MWe    & \cite{REF07, REF08} \\ \hline
HOMER                            &  USA                  &  Na                                         & UO$_2$       & Fast                     & 0.02 MWe        & \cite{POST-2000} \\ \hline
KRUSTY                           &  USA                  &  Na                                         & U-Mo            & Fast                     & 0.001 - 0.01 MWe & \cite{WN-2019, BRIGGS-2018} \\ \hline
Nominal OPUS concept   &  France              &  He-Xe                                   & UO$_2$        & Fast                     & 0.1 MWe           &\cite{REF11, REF12} \\ \hline 
NPPU                               &  Russia              &  He-Xe                                   & Unknown      & Unknown              & 1 MWe             & \cite{REF13} \\ \hline
SAFE-400                        &   USA                 & Na                                           & UN               & Fast                     & 0.1 MWe           & \cite{WN-2019} \\ \hline
SNAP-10A                        &  USA                  &  NaK                                       & U-ZrH$\rm{_x}$ & Thermal         & $<$ 0.001 MWe & \cite{WN-2019} \\ \hline
SP-100                             &  USA                  &  Li                                           & UN               &  Fast                     & 0.1 MWe          & \cite{REF03, REF04, REF05} \\ \hline
Topaz-1                            &  Russia               & NaK                                        & UO$_2$       & Thermal               &0.005 - 0.01 MWe& \cite{WN-2019} \\ \hline 
Topaz-2                            &  USA/Russia      & NaK                                        & UO$_2$       & Thermal                &0.006 MWe      & \cite{POLAN-1995} \\ \hline 
\end{tabular}
\end{adjustbox}
\label{Tab:SpaceReactors}
\end{table}

Historically, many space reactor designs, and also some modern designs, have employed Highly Enriched Uranium (HEU) fuel, with a number of designs employing fuel enriched above 90 wt.\% such as TOPAZ, SNAP-10A and KRUSTY \citep{GENK-2004, WN-2019}. Some reactor designs have used Low Enriched Uranium (LEU) fuel; however, there is an inherent trade-off with such designs: for the same power and length of reactor operation, the use of LEU fuel makes the reactor heavier and larger in dimension, resulting in significantly lower power densities \citep{LEE-2015}. Hence, for long duration missions and higher power outputs HEU fuel has been the preferred choice \citep{KHAND-2020}. Nevertheless, the obstacles with using HEU fuel should not be ignored, there will be additional security measures and associated costs with using HEU fuel \citep{NAM-2016}. The motivation for a high-power density, long-life system in this study results in the use of HEU material. The fuel enrichments studied here range between 20 and 90 wt.\%, noting that the original OPUS design employed fuel enriched to 93 wt.\% \citep{REF11, REF12}.

For terrestrial based energy systems employing gas-coolant, helium is the preferred coolant due to its high heat transfer coefficient \citep{PEAK-2018}. However, lighter coolants (such as He) result in high aerodynamic loading on the compressor, thereby requiring either a higher rotational speed or larger impeller size to compensate \citep{REF19}. The higher rotational speed or larger impeller size result in larger turbo-machinery and hence additional weight and/or volume. Therefore, for space applications, there is a compromise between choosing a coolant with a sufficiently high heat transfer coefficient and a heavier gaseous coolant (such as Xe). Therefore, a He-Xe mixture is preferred for space applications \citep{REF19}.

In this study, the nominal French OPUS design (0.1~MWe) was taken as a starting point, with this design chosen for further optimisation and investigation because:

\begin{itemize}
\itemsep-4pt
\item this core concept was considered to have the largest potential for optimisation;
\item the concept can operate at the high outlet temperatures required due to the use of a gaseous coolant and the selected materials (1300~K); and
\item the details of this concept are well documented in the open literature.
\end{itemize}

The key aims were to develop a space reactor with high-power output (around 1~MWe) and significantly higher power density than previous published gas-cooled reactor designs, which are in the 100~kWe range. Table \ref{Tab:KeyParams} summarises the key core parameters that the design in this study must meet. To ensure the launch mass and volume are minimised, several core optimisations have been proposed in this study to increase power density. A reduction in core volume will have neutronic and thermal implications, both of which have been modelled using the methods and modifications outlined in Section \ref{Sec:Method}. The modifications include: core geometry and fuel element changes necessary to enhance heat transfer from the graphite surface resulting in an increased power density; and radial enrichment zoning which has also been incorporated into the design to flatten the power profile across the core, leading to a reduction in peak operating temperature and a subsequent up-rate in power density.

\begin{table}[h!]
\centering
\caption{Key core parameters for the space reactor iterations}
\label{Tab:KeyParams}
\begin{tabular}{|c|c|c|}
\hline
\textbf{Parameter}     &  \textbf{Value} & \textbf{Reference(s)} \\ \hline
Effective Full Power Days Operation  & 2000 days &     \cite{REF11}             \\ \hline
Beginning of life excess reactivity  & 3000~pcm &     \cite{REF11}             \\ \hline
Peak Fuel Temperature  & 1900~K &    \cite{IAEA-2010, REF12}          \\ \hline
Inlet Temperature            & 880~K  &     \cite{REF12}         \\ \hline
Outlet Temperature         & 1300~K &    \cite{REF12}          \\ \hline
\end{tabular}
\end{table}

\section{Method}
\label{Sec:Method}

The method for designing suitable uprated reactor concepts starting from the nominal OPUS design was based on an iterative approach between neutronic modelling, using ERANOS 2.0 (see Section \ref{Sec:NeutModel}), and thermal-hydraulic performance using the thermal dimensioning tool outlined in Section \ref{sec:TDT}. At all points in the iterative process, the designs were screened against the key parameters highlighted in Table \ref{Tab:KeyParams}.

\subsection{The nominal OPUS core concept}

\begin{figure}[!h]
\centering
\includegraphics[angle=0,width=50mm]{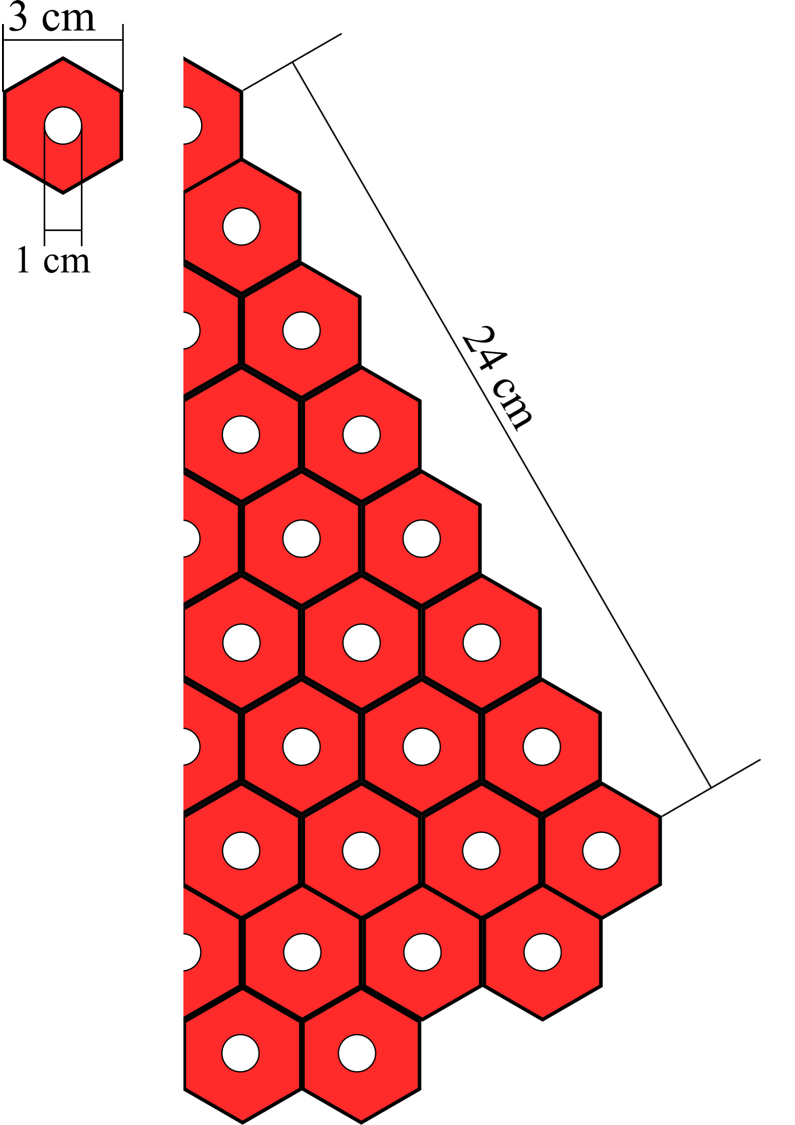}
\caption{The nominal OPUS core layout}
\label{Fig:SpaceReactorCores}
\end{figure}

The OPUS reactor concept \citep{REF11} is a gas-cooled fast reactor developed by CEA (see Figure \ref{Fig:SpaceReactorCores}). It is designed to cover a wide range of power outputs (100 to 500~kWe). However, it is only the 100~kWe variant that is described in sufficient detail in the open literature to develop a model of the core. No information was found on the thermal output of the 100~kWe system but based on the specific heat capacity ($C_p$) for coolant employed in OPUS design (which was 244.54~J/kg/K), the temperature rise across the core ($\Delta T$), which was 420~K and the reported mass flow rate ($ \dot{m}$), which was 3.6~kg/s \citep{REF12}, the thermal output is estimated to be around 370~kWth using Equation \ref{Eqn:Qdot}.

\begin{equation}
\dot{Q} = \Delta T \dot{m} C_p
\label{Eqn:Qdot}
\end{equation}
where $\dot{Q}$ is the rate of change of energy. Hence the thermal efficiency is determined to be 27\%.

The original core design contains 252 hexagonal fuel elements arranged to form the core cylinder. The core itself comprises 4 individual core quadrants. These are designed to separate in the event of launch failure in order to reduce the risk of inadvertent criticality if immersed in clay or water. A He-Xe coolant mix at a core inlet temperature of 880~K with an atomic mass of 85~g/mol enters the top of the core vessel, cooling the reactor pressure vessel. The He-Xe gas enters the bottom of the core and flows through coolant channels centrally located in every hexagonal fuel element. Ultimately, the coolant leaves the core with an average outlet temperature of 1300~K, driving a direct cycle Brayton engine. Direct cycle Brayton engines are the preferred energy conversion option for space reactor propulsion in the range 0.1~MWe to 10's~MWe since they allow for compact and low specific mass (kg/MWe) turbo-machinery, thereby aiding in reducing launch mass \citep{REF19}. The key parameters of the CEA original OPUS concept are summarised in Table \ref{TABLE01}.

The fuel elements in the nominal OPUS concept have a hexagonal pitch of 3 cm with a central 1 cm diameter coolant channel. The fuel comprises a graphite matrix with 45\% (by volume) bistructural isotropic (BISO) fuel particles \citep{REF11}. The kernels are UO\textsubscript{2} and enriched to 93~wt.\%.

\begin{table}[h!]
\centering
\caption{Nominal OPUS core key parameters \citep{REF11, REF12}.}
\label{TABLE01}
\begin{tabular}{|c|c|}
\hline
\textbf{Parameter}                                                               &   \textbf{Nominal OPUS Core}    \\ \hline
Thermal power                                                                    & 0.370~MWth                                 \\ \hline
Electrical power                                                                   & 0.100~MWe                                         \\ \hline
Active core diameter                                                           & 48 cm                             \\ \hline
Active core height                                                               & 48 cm                                           \\ \hline
$^{235}$U enrichment                                                        & 93 wt.\%                                          \\ \hline
\begin{tabular}[c]{@{}c@{}}Fuel element pitch\\ (across flats)\end{tabular}      & 3 cm                            \\ \hline
Coolant channel diameter                                                         & 1cm                                                   \\ \hline
Fuel elements in core                                                           & 252                         \\ \hline
Kernel type                                                                           & UO$_2$                                                                  \\ \hline
Coolant type                                                                     & 85 g/mol He-Xe                                      \\ \hline
Coolant flow rate                                                                & 3.6 kg/s                                                                         \\ \hline
Radial reflector thickness                                                   & 8 cm                                                                   \\ \hline   
Axial reflector                                                                     & 8 cm                                           \\ \hline
\end{tabular}
\end{table}

Note that the fuel employed in the reactor design developed here is the same type of BISO fuel that is proposed in the nominal OPUS design but with a uranium nitride (UN) kernel, as discussed in Section \ref{Sec:FuelKernelTemp}. BISO fuel was the first demonstrated successful coating technology with the name referring to the two material components (the fuel kernel and the PyC) and their isotopic geometries \citep{IAEA-2010, HOW-1978}. Further developments in coated particle fuel resulted in an additional layer of silicon carbide (SiC) and resulted in the so-called TRISO (tristructural isotropic) fuel concept \citep{IAEA-2010, HOW-1978}. The purpose of the additional SiC is to provide a high degree of mechanical and thermal resistance. Moreover, the SiC layer acts as a leak tight seal around the previous layers and is particularly good at retaining the metallic fission products and tritium \citep{IRSN-2015, SAWA-2012, VERF-2012}. Table \ref{Tab:TRISO_vs_BISO} provides a comparison between conventional TRISO fuel in modern High Temperature Reactor (HTR) designs and the BISO fuel employed in the OPUS and JANUS concepts studied here. It is important to note that in conventional TRISO fuel the kernel constitutes around 35\% of the total particle volume, whereas in the BISO particles used in this study the kernel make up around 70\% of the total particle volume. Therefore, with the BISO fuel concept, for a given packing fraction, there is a greater concentration of fissile material present in the core. Hence, whilst BISO fuel has been superseded in many applications by TRISO fuel, BISO fuel is still a popular choice for space reactors due to the higher volume fraction of fissile material \citep{REF11, REF12}. The higher fuel volume fraction is important when a compact and light-weight core design is required.

\begin{table}[h!]
\centering
\caption{Comparison between conventional TRISO particles \citep{PETTI-2012} and OPUS/JANUS BISO particles \citep{REF11}}
\label{Tab:TRISO_vs_BISO}
\begin{tabular}{|c|c|c|}
\hline
\textbf{Layer}         & \textbf{TRISO} & \textbf{BISO} \\ \hline
UO$_2$ kernel radius                    &  \SI{500}{\micro\meter} & \SI{820}{\micro\meter} \\ \hline
Porous carbon thickness                &  \SI{90}{\micro\meter}  & \SI{45}{\micro\meter} \\ \hline
Inner Pyrolytic Carbon thickness    &  \SI{40}{\micro\meter}  & - \\ \hline
SiC thickness                                  &  \SI{40}{\micro\meter}  & -  \\ \hline
Outer Pyrolytic Carbon thickness   &  \SI{40}{\micro\meter}  & \SI{45}{\micro\meter} \\ \hline
\end{tabular}
\end{table}

As noted, the drawback of adopting BISO fuel is the increased likelihood of fission product release relative to TRISO particles \citep{GULD-1977, PETTI-2012}. However, given that the reactor core will begin operation only once launch is complete \citep{BEN-2002} and unmanned missions are considered, as outlined in Section \ref{sec:intro}, the higher fission product release can be tolerated.

The volumetric packing fractions in the OPUS designs and variants considered here are 45\%. This is high by historical standards (with many HTR systems having packing fractions less than 35\% \citep{PETTI-2005}); however, packing fractions up to 50\% have been considered for prismatic fuel designs \citep{PETTI-2007}. There is general agreement that high packing fractions in prismatic designs result in an increased likelihood of fuel failure \citep{PETTI-2007}. It should be noted that whilst an increased likelihood of fission product release can be tolerated, given the overriding priority of a light-weight design, it is important that steps are taken to minimise fission product release where possible. This is because the high operating temperatures of the core and targeted long core-life mean that in the presence of a large release of chemically aggressive fission products, this could result in failure of components within the reactor or turbo-machinery.

The fuel particles considered here have \SI{800}{\micro\meter} fuel kernel diameter and \SI{100}{\micro\meter} combined thickness of porous and pyrolytic carbon \citep{KIN-2003}, hence each particle was around 1~mm in diameter. Note that particulate fuel typically has kernel diameters of around \SI{500}{\micro\meter}, with some limited experience with \SI{800}{\micro\meter} fuel kernels \citep{SIM-2002}.

For the nominal OPUS core, reactivity during normal operation is controlled by 12 external reflector vertically hinged shutters \citep{REF11}. By opening the shutters, neutron leakage can be varied, which for such a small core significantly impacts overall core reactivity. Under operational conditions and with all shutters closed, k-eff would be 1.04. With all shutters open, k-eff would be 0.94 \citep{REF12}.

\subsection{Neutronic modelling}
\label{Sec:NeutModel}

The BISTRO module within the ERANOS 2.0 code suite \citep{REF14} was used in conjunction with the JEF-2.2 Nuclear Data Library \citep{REF15} to model an R-Z representation of the core, reactor pressure vessel (RPV) and surrounding reflector regions \footnote{Whilst ERANOS 2.3 with JEFF-3.1 Nuclear Data Library could be considered state-of-the-art, for the purpose of this scoping study ERANOS 2.0 with JEF-2.2 is considered acceptable.}. In all ERANOS calculations the individual fuel elements (comprising central coolant channels, graphite and particle fuel) were modelled homogeneously. Furthermore, consistent with other Gas-cooled Fast Reactor studies \citep{PONYA-2017, CRUZ-2006}, the fuel elements were modelled isothermally, but in this case with a temperature of 1580~K.

Initially, there was some concern over homogenising the fuel elements at the start of the calculation sequence and that using an R-Z representation of the core in subsequent calculations would introduce a large discrepancy. In addition, the use of ERANOS (a fast reactor code) to model a system that contains a significant amount of  graphite and beryllium, and would be expected to have a softer spectrum than a typical fast reactor, also raised concerns around the validity of the approach adopted. Therefore, confirmatory Monte Carlo calculations were performed using SERPENT-2 (2.1.23) with the JEF-2.2 Nuclear Data Library \citep{REF16} to highlight the impact of homogenising the core design, as well as to validate the ERANOS calculation route. For the purposes of this confirmatory calculation, fuel temperatures of 1500~K were used in both the ERANOS and SERPENT calculations. It should be noted that SERPENT has cross-sections tabulated at specific temperatures (300~K, 600~K, 900~K, 1200~K and 1500~K). Note that the spectrum for the final core iteration developed in this study is included in Section \ref{lab:Summary}.

The results shown in Table \ref{TABLE02} indicate that homogenising the core has a relatively small impact on core reactivity (366~pcm). Furthermore, a difference of 654~pcm between the homogenous ERANOS calculation and a higher fidelity heterogeneous SERPENT calculation suggests the method used is sufficiently accurate for a scoping analysis of this nature.

\begin{table}[h!]
\centering
\caption{Reactivity difference between the homogeneous ERANOS model and the different SERPENT configurations}
\label{TABLE02}
\begin{tabular}{|c|c|}
\hline
\textbf{SERPENT Configuration}         & \textbf{Reactivity difference} \\ \hline
Homogeneous  & 288 $\pm$ 1 pcm                   \\ \hline
Heterogenous & 654 $\pm$ 1 pcm                   \\ \hline
\end{tabular}
\end{table}

The ERANOS model was used to assess the enrichments required in the core to maintain sufficient reactivity for at least 2000 effective full power days, and to develop radial enrichment zoning patterns that minimise power peaking throughout core life.  However, in order to predict peak temperatures in the core, and in doing so determine the size of the core, a bespoke thermal dimensioning computer code was developed. The code used fundamental heat transfer theory \citep{REF17} to quickly assess conditions in the coolant, graphite surface, through the graphite block and within the fuel particle itself, as outlined in the following section.

\subsection{Thermal dimensioning tool}
\label{sec:TDT}

\begin{figure}[!h]
\centering
\includegraphics[angle=0,width=150mm]{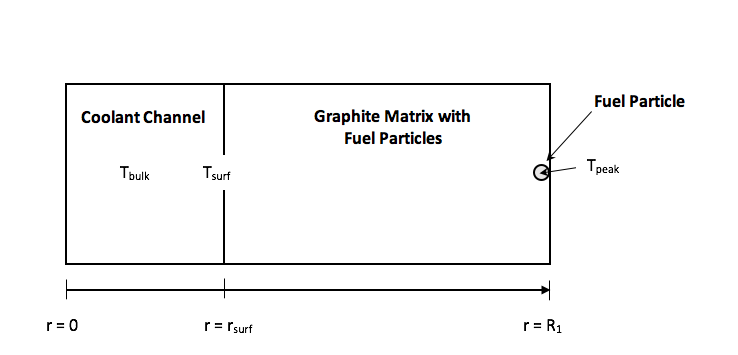}
\caption{Defined temperature regions within a particular axial zone, including the peak temperature (T$_{\rm{peak}}$) which is at the centre of fuel particle (not to scale) neighbouring the fuel element outer surface at r $=$ R$_1$. The centre of the coolant channel is located at r $=$ 0.}
\label{Fig:RadialSliceProblem}
\end{figure}

The calculation sequence (summarised in Figure \ref{Fig:RadialSliceProblem}) for determining the peak temperature (T$_{\rm{peak}}$) starts with determining the bulk temperature (T$_{\rm{bulk}}$) of the coolant, then the temperature at the surface of the fuel element (r = r$_{\rm{surf}}$) - where fuel element refers to the individual hexagonal structure with a central coolant channel and graphite matrix with fuel particles inside (see Figure \ref{Fig:RadialSliceProblem}). To simplify the method, the hexagonal fuel element has been modelled as a cylinder with the same cross sectional area. In actual fact, the temperature at the corner of the hexagonal fuel assembly will be slightly higher since the distance from the coolant channel will be greater than that of the cylindrical radius. However, as adiabatic boundary conditions have also been assumed, this will likely counteract this non-conservative assumption.

The calculation is followed by determining the temperature at the outer surface of the fuel element r = R$_1$ and finally the peak temperature, which is determined at the temperature at the centre of a fuel particle that neighbours the surface r = R$_1$. This process is captured in the thermal dimensioning tool, which for a particular core configuration can determine T$_{\rm{peak}}$ and was developed as part of this study. 

The thermal dimensioning tool works on the following basis:

\begin{itemize}
\item The temperature increase along the fuel element is determined using the power in each R-Z zone from the ERANOS model, along with the specific heat capacity and an estimate for the mass flow rate, to allow for the bulk coolant temperature at any location to be determined. The heat capacity employed for the nominal Xe-He coolant mixture was 20.8~J/mole/K \citep{REF19}.
\item Calculating the radial temperature increase from the bulk coolant to the surface of the graphite block as detailed in Section \ref{Sec:BulktoSurf}.
\item Calculating the radial temperature profile from the graphite surface to the fuel element outer surface as detailed in Section \ref{Sec:SurftoFuel}.
\item Calculating the temperature of a fuel kernel in closest proximity to  the fuel element outer surface as detailed in Section \ref{Sec:FuelKernelTemp}.
\end{itemize}

\subsubsection{Temperature increase from bulk coolant to the surface of the graphite block}
\label{Sec:BulktoSurf}

Starting with Equation \ref{Eqn:heat_tran_pA} for the heat transfer rate per unit area (W/m$^2$):
\begin{equation}
q''=h(T_{\rm{surf}} - T_{\rm{bulk}})
\label{Eqn:heat_tran_pA}
\end{equation}
where $h$ is the heat transfer coefficient, and $T_{\rm{surf}}$ and $T_{\rm{bulk}}$ are the temperatures at the positions shown in Figure \ref{Fig:RadialSliceProblem}. The heat transfer coefficient is given by

\begin{equation}
h = \frac{k_{\rm{mix}} \times N_{u}}{D_H}
\label{Eqn:h}
\end{equation}
where $k_{\rm{mix}}$ is the thermal conductivity of the particular coolant mixture, $N_{u}$ is the Nusselt number, which in this study it is determined through the Kays correlation \citep{REF20}, and ${D_H}$ is the equivalent diameter. 

Rearranging Equation  \ref{Eqn:heat_tran_pA} gives:

\begin{equation}
T_{\rm{surf}} = T_{\rm{bulk}} + \frac{q''}{h}
\label{Eqn:T_surf}
\end{equation}
where the heat transfer rate per unit area can be determined for a particular axial zone using the heat surface area and the power produced in each R-Z mesh from ERANOS.

\subsubsection{Temperature increase from the graphite surface to the fuel element outer surface}
\label{Sec:SurftoFuel}

Starting from the cylindrical form of the heat balance equation \citep{REF17} and assuming: 1) steady-state conditions; and 2) that thermal energy will primarily flow in the radial direction, the cylindrical form of the heat balance equation becomes:

\begin{equation}
\frac{1}{r}\frac{\partial \left(k r \frac {\partial{T}}{\partial{r}}\right)}{\partial r} = -\dot{q}
\end{equation}
where $\dot{q}$ is the energy generation per unit volume in the graphite block of the fuel element (W/m$^3$), which is again determined using the power density from ERANOS, $r$ is the radial distance through the fuel element (m), $k$ is the effective thermal conductivity of the graphite-fuel particle mixture (W/m/K) and $T$ is the temperature (K). Note that the effective thermal conductivity within the graphite block (BISO fuel particles and graphite) is determined using Maxwell's method detailed in \citep{CARS-1959}.

 Assuming no significant variation in thermal conductivity as a function of radius and applying the following boundary conditions: at the fuel element outer radius ($R_1$) no heat transfer is assumed (i.e. an adiabatic boundary condition); and the fuel temperature at the inner radius ($R_0$) is given by $T_{\rm{surf}}$, the temperature at a given radius within the graphite-fuel particle mixture is provided by Equation \ref{eqn:T_v_r}.
 
 \begin{equation}
 T = T_{\rm{surf}} + \frac{\dot{q}}{2k} \left [ \left ( \frac{R_0^2 - R_1^2}{2} \right ) + R_1^2 \ln \left ( \frac{R_1}{R_0} \right ) \right]
 \label{eqn:T_v_r}
 \end{equation}

\subsubsection{Temperature increase across a coated particle located on the fuel element outer surface}
\label{Sec:FuelKernelTemp}

The radial temperature distribution across the coated particle is determined via Fourier's Heat Equation and assuming the pyrolytic carbon thermal conductivity does not vary with radius, resulting in:

\begin{equation}
T(r) = T_0 + \frac{q}{4\pi k} \left ( \frac{1}{r} - \frac{1}{R_0} \right )
\end{equation}
where $q$ is the total heat transfer rate from a single particle, $k$ is the thermal conductivity across the carbon layer of interest, and $R_0$ relates to the radius at the boundary condition. $T_0$ and $R_0$ are initially set to the values relating the temperature and radius at the outer surface of the coated particle, respectively. $T(r)$ is used to calculate the radial temperature increase across the pyrolytic carbon layer and then, using this result as the new boundary condition ($T_0$),  the temperature increase across the porous carbon layer is calculated.

Finally, using the heat balance equation in spherical coordinates, the temperature at the centre of the coated particle (i.e. when $r = 0$) is calculated via:

\begin{equation}
T(r=0) = T_0 + \frac{{\dot{q} t_{\rm{kern}}}}{6 k} - \frac{ {\dot{q}} r^2} {6 k}
\end{equation}
where ${\dot{q}}$ is the energy per unit volume of an individual fuel kernel (determined using ERANOS), $k$ is the thermal conductivity of the fuel kernel and $t_{ \rm{kern} }$ is the kernel radius. The boundary condition, $T_0$, is determined using the previous calculation for the temperature increase across the porous carbon layer. According to \citep{REF24}, the thermal conductivity at 1800~K for UN is 26.2~W/m/K.  No data on irradiation degradation on thermal conductivity could be found so a lower value of 20~W/m/K was conservatively chosen. Note that as a consequence, the effective thermal conductivity of the fuel/graphite mix was calculated to be 20.44~W/m/K.




\subsection{Thermal modelling}

For all of the core iterations developed in this study, the R-Z ERANOS model comprised 100 radial and 13 axial zones making a total of 1300 equi-volume zones.  The thermal power per zone was predicted by ERANOS and used by the thermal dimensioning tool outlined in Section \ref{sec:TDT} to predict operating temperatures. For each of the 100 annuli, a 13-zone coolant channel surrounded by a single fuel element was modelled with a zero heat flux (adiabatic) outside boundary condition. The thermal dimensioning tool was used to:

\begin{itemize}
\itemsep-4pt
\item predict the temperature increase axially along the fuel stack; and
\item predict the temperature distribution radially from the bulk coolant, through the graphite-fuel matrix and a single fuel particle situated on the zero heat flux boundary.
\end{itemize}

\begin{figure}[!htbp]
  \centering
	\subfloat[]{\label{CEA_vs_NNL_v2}\includegraphics[width=0.46\textwidth]{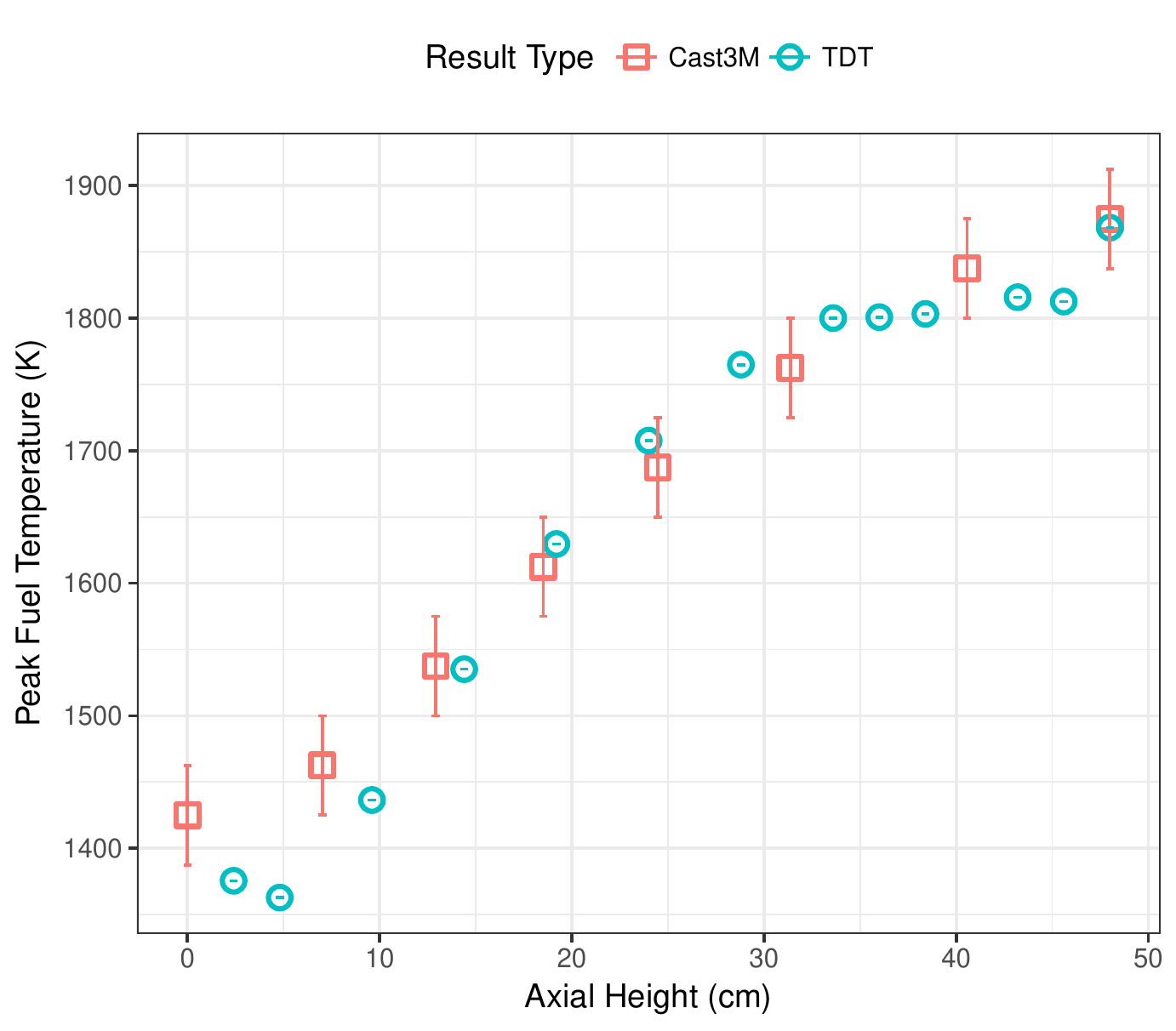}} 
  \hspace{0.2cm}               
  \subfloat[]{\label{NNL_temperature_power_data}\includegraphics[width=0.4\textwidth]{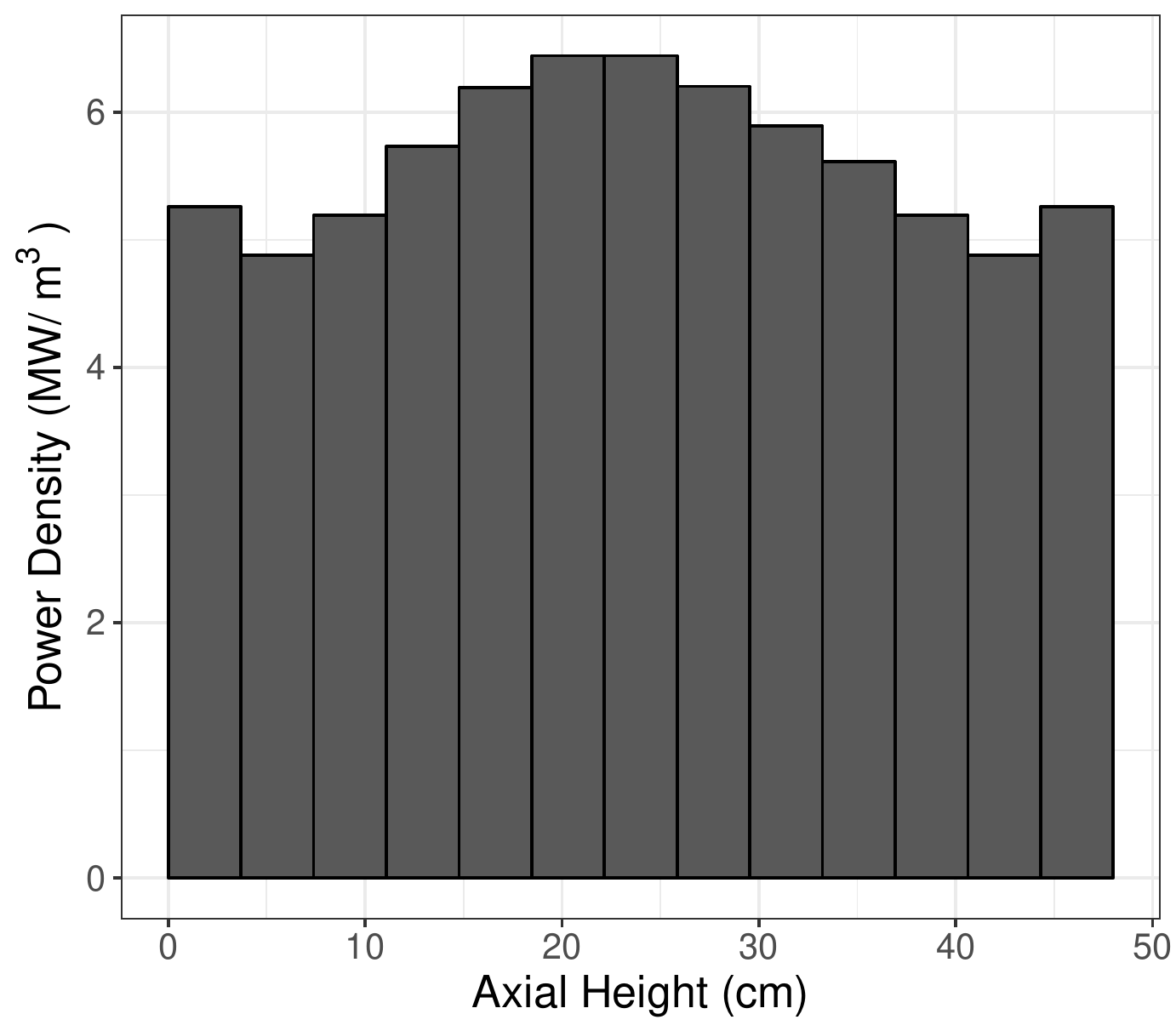}}
  \caption{Comparison of temperature distributions from CEA's Cast3M finite element software and the Thermal Dimensioning Tool (TDT) developed as part of this study (left) and calculated power distribution in each axial zone from ERANOS (right)}
  \label{Fig:CEA_vs_NNL}
\end{figure}

A temperature distribution for the nominal OPUS core concept with a given coolant flow rate and prescribed inlet and outlet coolant temperatures is available from \citep{REF12, REF22}. Using power densities from the ERANOS model of the nominal OPUS core developed in this study and the thermal dimensioning tool described above, the results from the thermal dimensioning tool was compared against the CEA results. The CEA thermal analysis employed the Cast3M finite element software package \citep{REF11}, which can provide temperature predictions of a higher accuracy than the thermal dimensioning tool developed in this study but with a higher computational cost. The code comparison was performed using the temperatures in the central fuel channel of the core where temperatures are highest and therefore most limiting. The results of the code comparison are shown in Figure \ref{Fig:CEA_vs_NNL}  

The error bars on the Cast3M results relate to the fact that Ref. \citep{REF11} gives the temperature distribution in each axial zone within a range of $\pm$37.5~K. As Figure \ref{Fig:CEA_vs_NNL}  shows, there is good agreement between thermal dimensioning tool and Cast3M. The code comparison performed between the TDT approach and Cast3M results shows that for determining peak temperatures the TDT approach is sufficient for the scoping and optimisation process undertaken in this study. However, some discrepancies would be expected between the TDT approach and Cast3M as a result of:


\begin{itemize}
\itemsep-4pt
\item The CEA calculation modelling irradiation induced graphite thermal conductivity degradation (the thermal dimensioning tool model assumes constant thermal conductivity throughout life); and
\item the conservative simplifications: ignoring radiative heat transfer from the core periphery and the (adiabatic) boundary conditions used to calculate the temperature increase through each fuel element.
\end{itemize}

Furthermore, the relatively flat fuel temperature profile towards the top of the core is due to the continual increase in enthalpy increase along the coolant channel balancing the reduction in power density produced within the fuel from one axial zone to the next (see Figure \ref{Fig:CEA_vs_NNL}). Towards the axial periphery of the core, the beryllium reflector will result in the reflection of more thermalised neutrons into the system, thereby resulting in elevated powers in these regions.

\section{Uprating the nominal OPUS design}

A key aim of this study was to develop a gas-cooled reactor for space applications (as outlined in Section \ref{sec:intro}) that can achieve electric power outputs close to 1~MWe. The nominal OPUS design already operates at close to the 1900~K summarised in Table \ref{Tab:KeyParams}, noting that the majority of the temperature increase from the bulk coolant to the peak fuel temperature is at the coolant channel surface. Hence, simply increasing the power output in the nominal core configuration will result in excessive temperature conditions. Therefore, all design variants included radial enrichment zoning, with between 4 and 5 radial zones chosen (noting that the the nominal OPUS core employed only a single radial enrichment) to keep the peak temperature below 1900~K. Furthermore, all core designs developed in this study implemented UN fuel kernels to reduce the fuel enrichment and allow greater flexibility in the range of radial enrichments to aid improvements in power profiles. An alternative approach, not studied here, would have been to keep the enrichment fixed at 93~wt.\% and vary the packing fraction to simulate variations in fissile concentration.

It should be noted that whilst Refs. \citep{IAEA-2010, REF12} indicate upper temperature limits on the 1900~K, experience with TRISO fuel operating at such high temperatures is limited. German experience for conventional (UO$_2$) TRISO fuel found failure rates of 1.65$\times$$10^{-5}$ during heating tests up to 1900~K, with tests running for $\sim$10$^2$ hours \citep{VERF-2012}. Therefore, greater levels of testing would need to be performed to determine failure rates -- noting that failure rates are highly temperature dependent and may require some trade-off between targetted power density and acceptable failure rates for the purpose of reactor design developed here.

Many UN studies assume isotopically pure $^{15}$N fuel (between 95\% and 99.9\% $^{15}$N content) due to the parasitic absorption of the predominant $^{14}$N nuclide (which constitutes 99.6\% of naturally occurring nitrogen, with remaining 0.4\% being $^{15}$N) and also the formation of radioactive $^{14}$C during irradiation of $^{14}$N, which represents a disposal concern \citep{IAEA-2019, RAY-2012}. In the analysis performed here, the UN fuel is modelled as 100\% enriched $^{15}$N. A comparison on the reactivity impact of employing 95\% vs 100\% $^{15}$N is included in Section \ref{lab:Summary}.

Note that the use of a high-density UN fuel operating at temperatures up to 1900~K in coated particle form would need to be validated. There currently appears to be limited to no experience with irradiation of UN coated fuel; however, UN coated particles have successfully been fabricated \citep{LED-1992, LED-1996, GANG-1993}. Overall, the extent of irradiation experience with UN is far lower than UO$_2$ under fast reactor conditions, with UN exhibiting low creep rates and high swelling rates, which may prove limiting for this design \citep{GUER-2012}. An experimental programme would be required to qualify the fuel form proposed in this study (i.e. BISO fuel with a UN kernel operating at high temperatures with relatively high packing fractions).

In addition to UN kernels and radial enrichment zoning, optimisation of the higher power output core configuration focused on:
\begin{enumerate}
\item Increasing the core size. This allowed an increase in power density to 10.7~MW/m$^3$ compared with the nominal OPUS design, which has a power density of  4.3~MW/m$^3$. 
\item Investigating the impact of changing the aspect ratio of the core (e.g. taller core configurations).
\item Increasing the number of fuel elements and reducing the coolant channel size, whilst keeping the total coolant channel cross section within the core constant. 
\item Modifying the coolant gas composition from 85~g/mol He-Xe to 20~g/mol.
\end{enumerate}

As will be detailed, whilst in principle items 2 and 4 above can allow for significant improvements in the power density, the inevitable engineering trade-offs associated with gas-cooled reactors are ultimately too penalising to make these options viable. Option 3 (resulting in the design iteration referred to as JANUS) successfully resulted in a much higher power density than the nominal OPUS core (4.3~MW/m$^3$ vs 26.5~MW/m$^3$) without unacceptable engineering trade-offs.

Regarding option 4, a reduction in coolant atomic mass from 85~g/mol to 20~g/mol permitted a 34\% reduction in core volume. However, the higher helium content will adversely impact the volume and/or mass of turbo-machinery required to convert the thermal heat to electricity \citep{REF19}. Considering the reactor core is a relatively small fraction of the overall energy system mass (around one-third of total mass \citep{CLIQ-2011}), the 34\% reduction in core volume and mass was not judged by the authors to be significant enough to overcome the much larger turbo-machinery volume and mass needed, which will prohibitively add to the launch mass.

\subsection{Increasing core size and implementing radial enrichment zoning}
\label{sec:LargeCore}

Using the nominal OPUS core design as a starting point, the core thermal power was increased to 3~MWth and the core size increased in both radial and axial directions. In addition, the enrichment was modified such that there was sufficient excess reactivity at beginning of life (BOL) of around 3000~pcm to allow for 2000~Effective Full Power Days (EFPD) of operation, outlined in Table \ref{Tab:KeyParams}. Furthermore, radial enrichments were varied manually in order to flatten the power shape and therefore reduce peak operating temperature.  The thermal dimensioning tool outlined in Section \ref{sec:TDT} was then used to estimate the peak operating temperature for a core with a 880~K core inlet temperature and 1300~K core outlet coolant temperature and to ensure the radial enrichment zoning was sufficient to maintain peak fuel operating temperatures below 1900~K. 

To meet the criteria regarding excess reactivity at BOL and peak fuel temperatures, only 4 radial enrichments (ranging from 30~wt.\% in the core centre to 60~wt.\% on the core periphery, equivalent to a volume average enrichment of 50 wt.\%) were necessary, as outlined in Table \ref{Tab:LargeCore}. The final core design had a power density of 10.7 MW/m\textsuperscript{3} (compared with 4.3 MW/m\textsuperscript{3} for the nominal OPUS design). 

\begin{table}[h!]
\centering
\caption{Radial enrichment zones, starting from the centre of the core, for the core configuration with a radius of 35.5~cm and hexagonal fuel element pitch of 3~cm.}
\label{Tab:LargeCore}
\begin{tabular}{|c|c|c|}
\hline
\textbf{Radial zone number}         & \textbf{$^{235}$U enrichment} & \textbf{Proportion of core at this enrichment} \\ \hline
Zone 1  &  30 wt.\% & 10\% \\ \hline
Zone 2  &  40 wt.\% & 30\% \\ \hline
Zone 3  &  50 wt.\% & 10\% \\ \hline
Zone 4  &  60 wt.\% & 50\% \\ \hline
\end{tabular}
\end{table}

\subsection{Decreasing the core aspect ratio}

A number of cores were designed with aspect ratios ranging from 0.3 up to 1.5, where the aspect ratio is defined to be 2R\textsubscript{c}/H, where H is the active core height and R\textsubscript{c} is the active core radius. The core configuration with an aspect ratio of 1.0 is the same as that outlined in Section \ref{sec:LargeCore}. For each core model, radial enrichment zoning, as well as UN BISO fuel were used along with an the requirement for an excess reactivity of 3000~pcm at BOL. Furthermore, the fuel element geometry was the same as the nominal OPUS fuel design (3~cm hexagonal pitch with a 1~cm diameter central coolant channel) but with varying axial heights. For each case, the core volume was varied until the predicted peak temperature was around 1850~K, which is less than the peak operating temperature limit of 1900~K.

{\footnotesize
\tabcolsep=0.11cm
\begin{table}[h!]
\centering
\caption{Impact of adjusting core aspect ratio on peak temperature, power density and core pressure drop}
\begin{tabular}{|c|c|c|c|c|c|}
\hline 
\bf{Aspect} & \bf {Active} & \bf{Core}                             & \bf{Peak Fuel}           & \bf{Core Pressure} & \bf{Power Density}\\
\bf{Ratio}    & \bf {Height (m)} & \bf{Volume (m$^3$)}  & \bf{Temperature (K)} & \bf{Drop (MPa)}      & \bf{(MW/m$^3$)}\\ \hline
0.30 & 1.19 & 0.12 & 1840 K & 1.240 & 25.2 \\ \hline
0.50 & 0.94 & 0.16 & 1800 K & 0.365 & 18.2 \\ \hline
0.70 & 0.80 & 0.19 & 1840 K & 0.175 & 15.5 \\ \hline
1.00 & 0.71 & 0.28 & 1830 K & 0.064 & 10.7 \\ \hline
1.20 & 0.65 & 0.32 & 1820 K & 0.042 & 9.5  \\ \hline
1.50 & 0.61 & 0.40 & 1820 K & 0.022& 7.5  \\ \hline
\end{tabular}
\label{Tab:AspectRatios}
\end{table}
}

Table \ref{Tab:AspectRatios} shows the predicted peak operating temperatures along with the total core volumes and predicted pressure drops across the core. The results show that a factor of two increase in power density is achievable for a highly elongated core (i.e. aspect ratio of 0.3). However, the pressure drop across the core is over 1~MPa, similar to the operating pressure of the OPUS core. Such a design would require a significant increase in compressor size and additional structural support, probably resulting in no overall mass or volume savings, noting that usually only around a third of the power system mass is the reactor core itself \citep{CLIQ-2011}.

\subsection{Increasing the number of fuel elements and reducing the coolant channel size}

A number of different sized cores were modelled using ERANOS ranging from a relative core volume of 0.3 up to 1.5. For each core:

\begin{itemize}
\itemsep-4pt
\item the total power output was set to 3~MWth;
\item radial enrichment zoning was used to limit peak to average power production; and
\item an excess reactivity target at BOL of 3000~pcm was applied to ensure sufficient margin for at least 2000 EFPD.
\end{itemize}
The thermal dimensioning tool was then used to assess how large each fuel element would need to be in to keep the peak fuel temperature within the given limit.

As the fuel element hexagonal pitch is decreased, which results in the heat transfer path between the fuel and coolant reducing and a corresponding improvement in heat transfer. The core volume can therefore be reduced, as shown in Table \ref{Tab:FuelElementGeoms}. However, the reduction in coolant channel diameter and fuel element pitch results in an increase in pressure drop across the reactor core leading to the requirement for a larger and therefore heavier gas compressor.

Assuming a system pressure of 1.4~MPa, a relative surface roughness \SI{2}{\micro\meter} and a combined K-factor of 1.5, estimates for the pressure drop \citep{REF17} across the core have been calculated and are given in Table \ref{Tab:FuelElementGeoms}. Although large for the smaller core designs, the pressure drop across the turbine is likely to be much higher (we estimate using engineering judgment this to be around half the core system pressure (i.e. 0.7~MPa)). The core configurations summarised in Table \ref{Tab:FuelElementGeoms} were based on maintaining a core aspect ratio of 1.0 (i.e. core height $=$ core diameter); fuel elements with equivalent coolant channel to fuel matrix area ratios; and employing the thermal dimensioning tool outlined in Section \ref{sec:TDT} with a target peak fuel temperature of approximately 1850~K.

As shown in Table \ref{Tab:FuelElementGeoms}, reducing fuel element pitch will lead to an increase in core pressure drop and therefore an increase in compressor size and volume in order to pump the coolant through a more restricted core. However, for all but the smallest fuel element design, the increase in compressor mass and volume required to increase reactor inlet pressure head will not be significant and will potentially be offset by the large reduction in core volume as a result of improved heat transfer. Therefore, an increase in the number and a reduction in individual coolant channel surface area is considered to be very effective at increasing core power density.

To ensure the use of homogeneous ERANOS models for cores significantly larger than the nominal OPUS design was still broadly valid, a comparison was performed on one of the larger core configurations, namely the core with a radius of 35.5~cm and 71.0~cm active height outlined in Table \ref{Tab:FuelElementGeoms}. The discrepancy between the homogeneous ERANOS model vs the SERPENT model for this larger core was found to be 333~$\pm$~1~pcm, which is similar to the 288~pcm discrepancy observed when modelling the smaller nominal OPUS design (see Section \ref{Sec:NeutModel}).

\begin{table}[h!]
\centering
\caption{Effect of changing fuel element geometry on power density and core pressure drop when maintaining a core aspect ratio of 1.0 (i.e. core height $=$ core diameter)}
\begin{adjustbox}{width=1\textwidth}
\small
\begin{tabular}{|c|c|c|c|c|c|c|}
\hline
\bf{Coolant channel} & \bf{Distance between}  & \bf{Active core} & \bf{Active core}  & \bf{Power}               &  \bf{Peak fuel}      & \bf{Core}  \\ 
\bf{diameter}             & \bf{flats of hexagonal}   & \bf{height}         & \bf{radius}          & \bf{density}             & \bf{temperature}  & \bf{pressure drop} \\ 
\bf{(mm)}                   & \bf{fuel element (mm)} & \bf{(cm)}            & \bf{(cm)}            &  \bf{(MW/m$^3$)}    & \bf{(K)}                 & \bf{(MPa)} \\ \hline
7.1                             & 21.5                              & 49.2                  & 24.6                  & 32.0                         & 1856                    & 0.27 \\ \hline
7.4                             & 22.3                              & 54.2                  & 27.1                  & 24.0                         & 1813                    & 0.19 \\ \hline
8.5                             & 25.7                              & 58.4                  & 29.2                  & 19.2                         & 1831                    & 0.14 \\ \hline
9.0                             & 27.2                              & 62.0                  & 31.0                  & 16.0                         & 1819                    & 0.11 \\ \hline
10.0                           & 30.0                              & 71.0                  & 35.5                  & 10.7                         & 1826                    & 0.06 \\ \hline
12.4                           & 37.2                              & 84.1                  & 42.1                  & 6.42                         & 1838                    & 0.03 \\ \hline
\end{tabular}
\end{adjustbox}
\label{Tab:FuelElementGeoms}
\end{table}

We assume a limit of 0.20~MPa for the pressure drop limit across the core so that the pressure drop across the core remains small compared with the pressure drop expected across the rest of the direct cycle loop. Hence the optimal core configuration from Table \ref{Tab:FuelElementGeoms} is the iteration with a coolant channel diameter of 7.4~mm and a power density of 24.0~MW/m$^3$. This iteration is referred to as the JANUS concept.

Note that for the JANUS concept, the minimum width region is 11.15~mm - 3.7~mm = 7.45~mm. Given the size of the particles (1~mm), the minimum width of the fuelled region is not much larger than the size of the particles (7.45 particles) means that the particles are not completely free to distribute in a random fashion. Thus, some of the particles will likely be distributed as quasi-crystals, and a significant fraction of particles will likely be touching each other. Hence, there is a significant risk of cracking and future work would need to investigate the likelihood of fuel failure, potentially necessitating irradiation trials. Further optimisation of the fuel configuration may be required to ensure a fuel failure fraction that is acceptable for the aims of the JANUS concept (noting earlier discussion on acceptability of higher failure fracture compared to conventional TRISO particles for this particular space reactor concept).

\subsection{Summary of optimisation choices and JANUS core characteristics}
\label{lab:Summary}

\begin{figure}[htbp]
\centering
\includegraphics[angle=0,width=115mm]{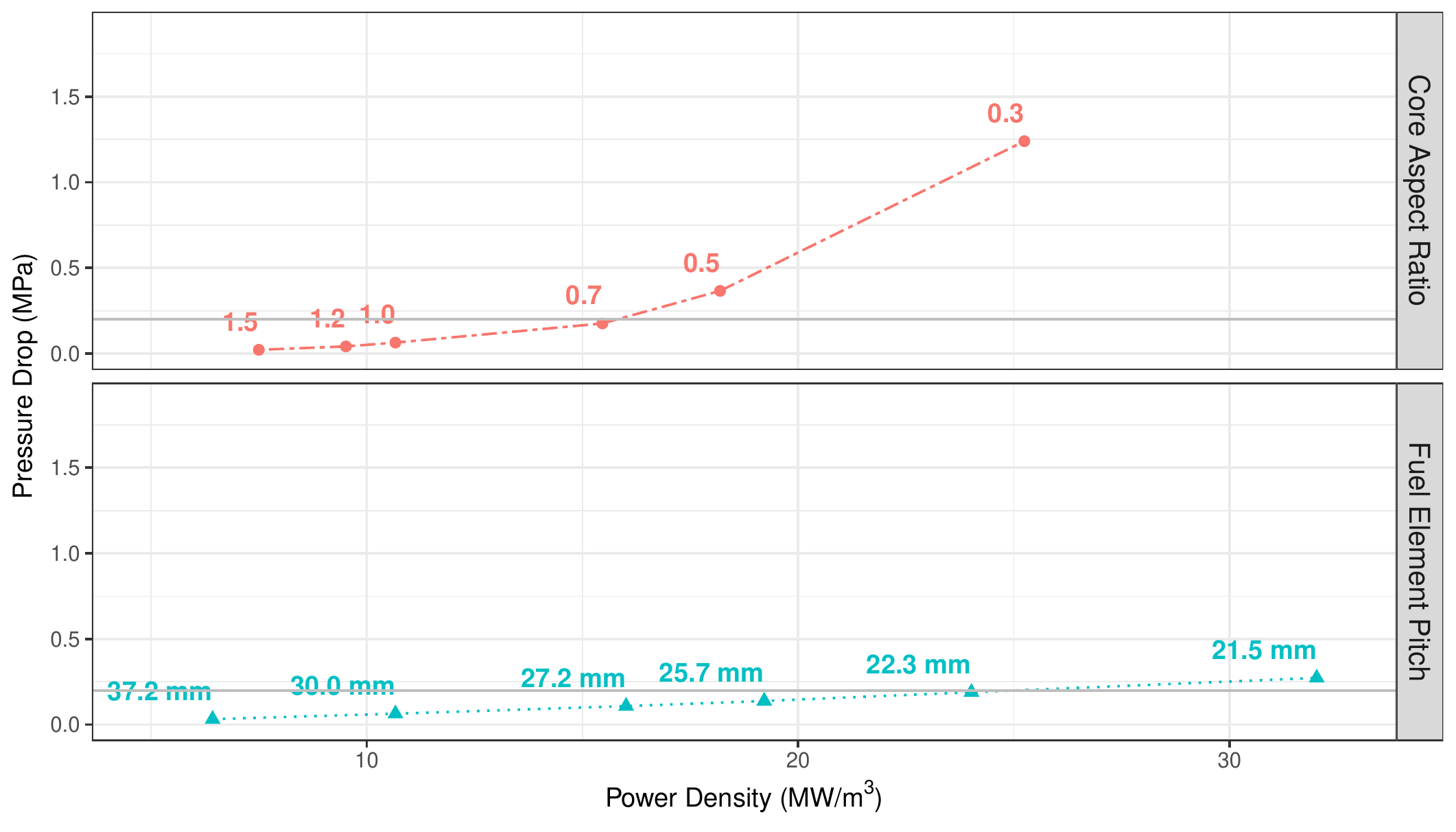}
\caption{The impact on pressure drop for two core design configurations as a function of power density, including the 0.2 MPa core pressure drop limit imposed on the core iterations}
\label{FIGURE_03}
\end{figure}

\begin{figure}[!h]
\centering
\includegraphics[angle=0,width=150mm]{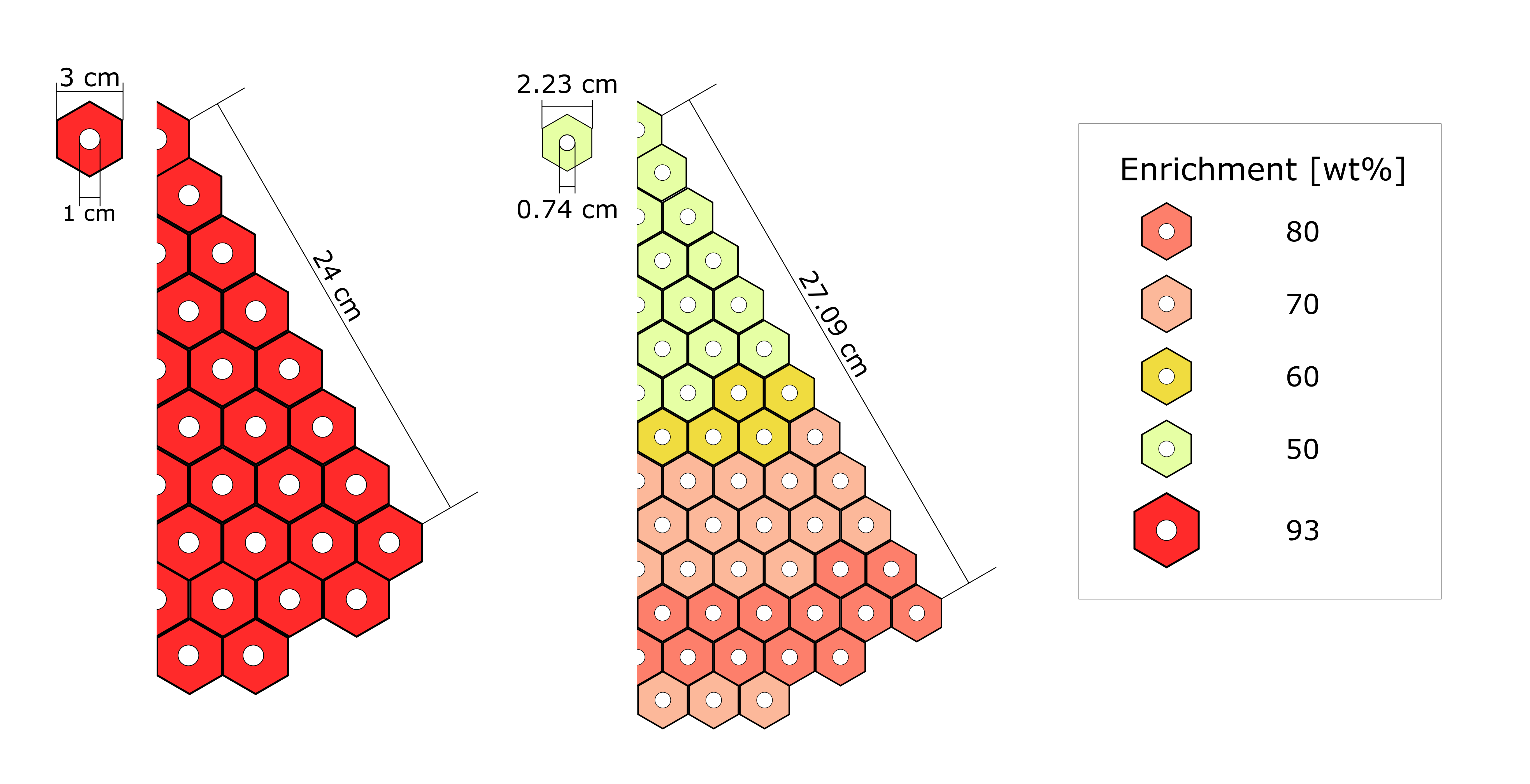}
\caption{Comparison of the nominal OPUS core and the JANUS variant developed in this study}
\label{Fig:SpaceReactorCoresv2}
\end{figure}

\begin{table}[h!]
\centering
\caption{Nominal OPUS core and the JANUS core key parameters}
\label{Tab:CoreComp}
\begin{tabular}{|c|c|c|}
\hline
\textbf{Parameter}                                                               &   \textbf{Nominal OPUS Core}  & \textbf{JANUS Core}  \\ \hline
Thermal power                                                                    & 0.370~MWth                                  & 3~MWth    \\ \hline
Electrical power                                                                   & 0.100~MWe                                   & 0.811~MWe                     \\ \hline
Active core diameter                                                           & 48 cm                                        & 54.2 cm          \\ \hline
Active core height                                                               & 48 cm                                        & 54.2 cm           \\ \hline
Mean $^{235}$U enrichment                                                        & 93 wt.\%                                    &  68 wt.\%                    \\ \hline
\begin{tabular}[c]{@{}c@{}}Fuel element pitch\\ (across flats)\end{tabular}      & 3 cm           &  2.23 cm                     \\ \hline
Coolant channel diameter                                                         & 1cm                                     &  0.74 cm                     \\ \hline
Fuel elements in core                                                           & 252          &  418                     \\ \hline
Fuel form                                                                              & \multicolumn{2}{c|}{\begin{tabular}[c]{@{}c@{}}Graphite matrix containing 45\%\\ BISO fuel particles by volume\end{tabular}} \\ \hline
Kernel type                                                                           & UO$_2$                                   & UN                                                      \\ \hline
Coolant type                                                                     & 85 g/mol He-Xe  &                 85 g/mol He-Xe                            \\ \hline
Coolant flow rate                                                                & 3.6 kg/s                                      &    29.2 kg/s                                                     \\ \hline
Radial reflector thickness                                                   & 8 cm                                           &  8 cm                                                    \\ \hline   
Axial reflector                                                                     & 8 cm                                           &  8 cm               \\ \hline
Number of radial enrichment zones                                   & 1             &    5 \\ \hline
Power density                                                                     & 4.3 MW/m$^3$                                           &  24 MW/m$^3$               \\ \hline
\end{tabular}
\end{table}

Simply increasing the core size and employing radial enrichment heterogeneity allows for an increase in power density from 4.3~MW/m$^3$ (the nominal OPUS design) to 10.7~MW/m$^3$, noting there will be a drawback associated with the larger coolant inventory and therefore larger turbo-machinery. For a given pressure drop sacrifice, reducing the fuel element pitch is far more effective at reducing core mass and volume as shown in Figure \ref{FIGURE_03}.  We assume a limit of 0.20~MPa for the pressure drop limit across the core so that the pressure drop across the core remains small compared with the pressure drop expected across the rest of the direct cycle loop. Elongating the core allows for a power density of approximately 16~MW/m\textsuperscript{3}, whilst ensuring the pressure drop is below 0.20~MPa. Conversely, reducing the fuel element pitch (whilst keeping the total coolant channel cross sectional area in the core constant), allows for a power density of 24~MW/m\textsuperscript{3}.

Table \ref{Tab:CoreComp} gives a comparison of the nominal OPUS core parameters \citep{REF11, REF12} and the JANUS core parameters. For completeness, a comparison of the two cores is included in Figure \ref{Fig:SpaceReactorCoresv2}.

\begin{figure}[!htp]
\centering
\includegraphics[angle=0,width=125mm]{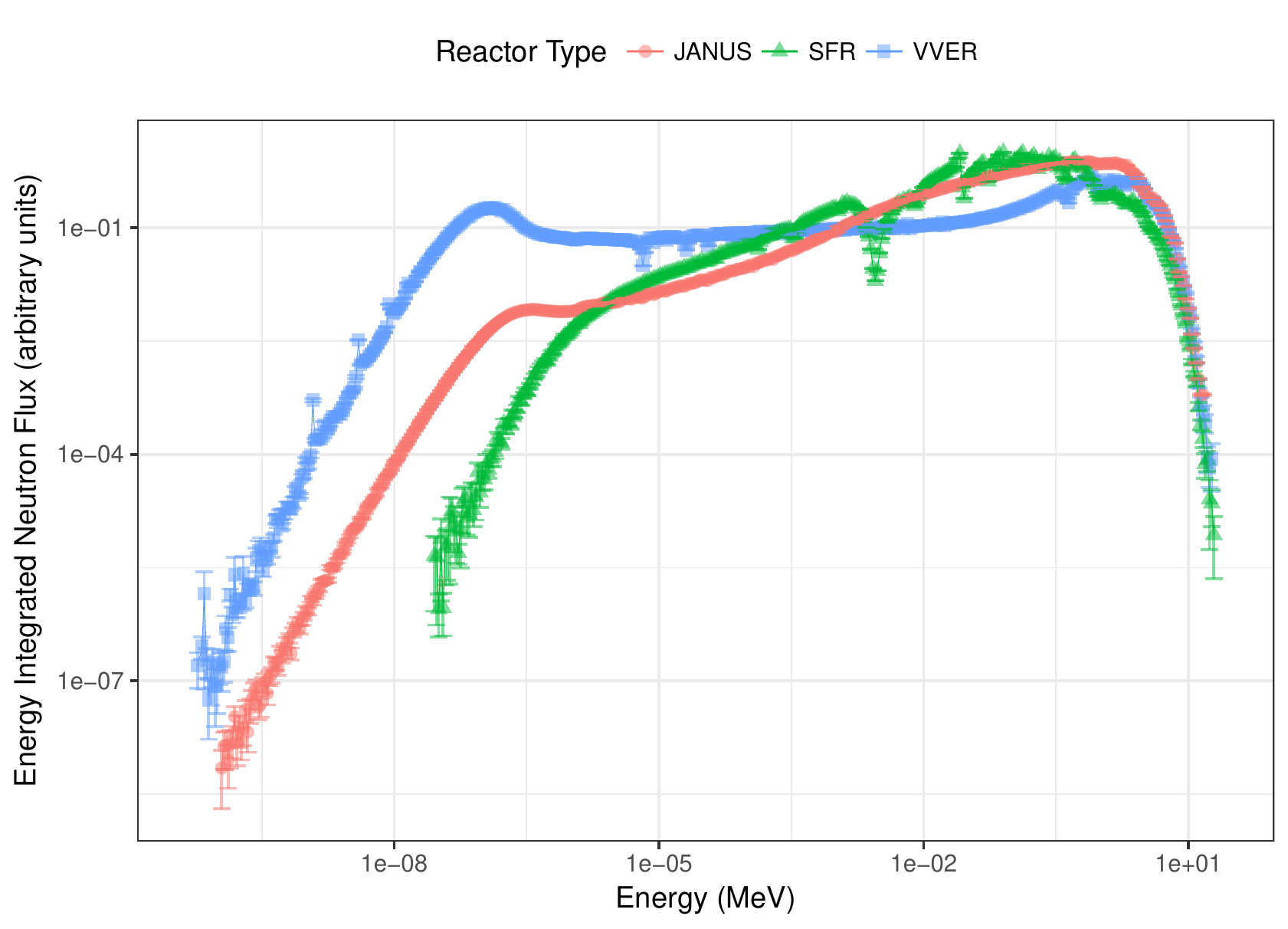}
\caption{Comparison of flux spectrum between JANUS, SFR and PWR (VVER) SERPENT models.}
\label{Fig:Flux_comparison}
\end{figure}

As a further comparison, the flux spectrum from the JANUS heterogeneous SERPENT model, taken across the whole-core including the beryllium reflector was compared with a typical SFR spectrum \citep{SCIO-2011} and PWR spectrum \citep{SERP-2020} as shown in Figure \ref{Fig:Flux_comparison}. The PWR spectrum is based on a  VVER-440 assembly design \citep{SERP-2020}. Note that the energy integrated flux has been scaled in all cases to a maximum of 1.0. Figure \ref{Fig:Flux_comparison} shows that the JANUS core exhibits a spectrum somewhere between the PWR and SFR cases modelled since, whilst a beryllium reflector is employed and there is a significant amount of graphite, this is offset by the high enrichment, the very high packing fraction of the BISO particles in the fuel and the use of a non-moderating coolant.

Finally, the impact of employing 95\% vs 100\% enriched fuel $^{15}$N was investigated for the JANUS model. Starting with the SERPENT heterogeneous model and performing the calculation at beginning-of-life, this showed a reactivity penalty of 79.2 $\pm$ 8.6 pcm.

\section{Future Work}

Whilst the design constraints employed here are consistent with those in the literature, it is important to note that peak temperatures of 1900~K, even for coated particle fuel, is high and operational experience of particles operating at such temperatures (even if the peak temperatures close to 1900~K are only sustained for a portion of the reactor's life) is only $\sim$10$^2$ hours. Furthermore, whilst there is experience with high packing fractions, the use of thin fuel elements proposed in the JANUS design may result in some of the particles distributed as quasi-crystals and therefore they may be in close contact, resulting in elevated failure probabilities. Hence, even though higher failure fracture than conventional HTRs can be tolerated for an unmanned mission, it would be important to ensure that failure rates are not onerously high; thereby allowing for large ingress of chemically aggressive fission products escaping the core and damaging system components. Future work would need to consider the trade-off between high power densities, high packing fractions and fuel failure rates. This would potentially necessitate irradiation trials for the fuel configuration proposed.

\section{Conclusions}


Many current and published gas-cooled space reactor designs operate at relatively low power output and low power densities, such as the CEA's OPUS concept, which employs UO$_2$ BISO fuel with a He-Xe gas coolant and has a power output of 0.100~MWe, with a power density of 4.3~MW/m$^{3}$. However, such low power outputs make these cores unsuitable for future space applications, including cargo support for a manned mission to Mars and outer Solar System missions with large mission payloads. Hence, there is a desire to develop a space reactor with high power density (to minimise launch mass and volume) and high power output (around 1~MWe).

Starting from this published OPUS design, a gas-cooled space reactor concept, referred to as JANUS, has been developed at the conceptual level, that operates at a much higher power output (0.8~MWe) and significantly increased power density 24~MW/m$^{3}$.  The fuel form adopted in the JANUS concept is uranium nitride BISO fuel to help reduce demands on enrichment and provide greater flexibility in radial enrichment zoning. The optimisation route to develop a viable gas-cooled space reactor involved neutronically and thermally modelling the different core designs to ensure key criteria related to core-life (2000~EFPD) and peak fuel operating temperatures (1900~K) were obeyed. In addition, comparisons involving Monte Carlo neutronic calculations using SERPENT were used to validate the ERANOS calculation route adopted in this study. The discrepancies between ERANOS and SERPENT of around 500~pcm showed that the approach adopted in this study was sufficiently accurate for the purpose of scoping out viable core configurations as part of the optimisation route. 

For thermal modelling, a simple yet robust thermal dimensioning tool has been developed as part of this study to ensure accurate but computationally efficient predictions on peak temperature so that the fuel could be expected to operate within its design constraints. The thermal dimensioning tool was shown to accurately predict peak fuel temperatures in the original OPUS design, when compared with the results from the CEA's Cast3M finite element calculations, and greatly aided in the optimisation process when considering many core variants.


A key outcome of this study related to identifying the inherent challenges with increasing power density when employing a gas-coolant operating with a direct Brayton cycle. The use of a gas-coolant inherently limits the achievable power density and the mechanisms to improve heat transfer between the coolant and fuel (to ensure fuel operating temperature are below the 1900~K limit) often result in larger pressure drops across the core and/or greater demands on the turbo-machinery. This resuls in unacceptable increases in launch mass and volume. Nevertheless, adopting a new fuel element design, that reduced the heat transfer path between the fuel and coolant, and employing greater radial heterogeneity has resulted in a core design capable of meeting the design constraints.

\section*{Acknowledgements}

This work was supported by the National Nuclear Laboratory's (NNL's) Innovation Research and Development fund. The authors would also like to gratefully acknowledge input received from Thomas Bennett (NNL), Bruno Merk (University of Liverpool) and Richard Stainsby (Wood plc).

\bibliography{References}

\end{document}